\documentclass[%
 reprint,
 amsmath,amssymb,
 aps,
prc,
]{revtex4-1}

\usepackage{graphicx}% Include figure files
\usepackage{dcolumn}% Align table columns on decimal point
\usepackage{bm}% bold math
\begin{document}

%\preprint{APS/123-QED}

\title{A Study of the Di-Hadron Angular Correlation Function in Event by Event Ideal Hydrodynamics}% Force line breaks with \\

\author{R.~P.~G.~Andrade}
\email{rone.andrade@gmail.com}
\author{J.~Noronha}
\email{noronha@if.usp.br}
\affiliation{Instituto de F\'{\i}sica, Universidade de S\~{a}o Paulo, C.P. 66318,
05315-970 S\~{a}o Paulo, SP, Brazil}

\date{\today}% It is always \today, today,

\begin{abstract}
The di-hadron angular correlation function is computed within boost invariant, ideal hydrodynamics for Au+Au collisions at $\sqrt{s}_{NN}=200$ GeV using Monte Carlo Glauber fluctuating initial conditions. When $0<p_T< 3$ GeV, the intensity of the flow components and their phases, $\left\{v_n, \Psi_n \right \}$ ($n=2,3$), are found to be correlated on an event by event basis to the initial condition geometrical parameters $\left\{\varepsilon_{2,n}, \Phi_{2,n} \right \}$, respectively. Moreover, the fluctuation of the relative phase between trigger and associated particles, $\Delta_n =\Psi_n^t - \Psi_n^a$, is found to affect the di-hadron angular correlation function when different intervals of transverse momentum are used to define the trigger and the associated hadrons. 
\end{abstract}

\maketitle

\section{\label{sec:introduction} Introduction}

The nontrivial structures in di-hadron angular correlation measurements with respect to a single charged (or neutral) high-$p_T$ trigger observed in heavy ion collisions \cite{Adams:2005ph,Adare:2008ae,Abelev:2009af,Chatrchyan:2011eka} are among the most important probes of the hot and dense matter created in these reactions. In fact, angular correlations measured in Au+Au collisions at RHIC with the center-of-mass energy per nucleon pair $\sqrt{s}_{NN}$= 200 GeV are significantly different than those observed in pp or d+Au collisions (this difference, however, seems to disappear for the higher energy collisions performed at the LHC \cite{Khachatryan:2010gv,CMS:2012qk}). In the longitudinal direction, the di-hadron correlation function is characterized by a long range structure in the relative pseudo-rapidity coordinate denominated ``ridge" \cite{Abelev:2009af} while in the azimuthal direction one finds three prominent peaks: the near side peak $\Delta \phi=0$ aligned with the trigger hadron and two other away side peaks that are symmetrically positioned with respect to $\Delta\phi=\pi$. This azimuthal profile indicates the existence of a considerable fraction of higher order harmonic flows, mainly triangular and quadrangular flows, in addition to the well known direct and elliptic flows. 

These angular correlations have been studied in the past in the context of the energy deposited by jets in a smooth hydrodynamic medium \cite{Stoecker:2004qu,CasalderreySolana:2004qm,Satarov:2005mv,Chaudhuri:2005vc,Renk:2008km,Noronha:2008un,Betz:2008wy,Betz:2008ka,Torrieri:2009mv,Betz:2010qh}. In \cite{Andrade:2008xh,Andrade:2006yh,Takahashi:2009na} it was suggested that the higher order harmonic flows are connected to the fluctuations in the initial conditions for hydrodynamics. In Ref. \cite{Takahashi:2009na}, it was shown that the fluctuations in the initial conditions, characterized by longitudinal tube-like structures, can in fact produce after hydrodynamic expansion the structures observed in the data. 

Considerable effort has been since given towards understanding how harmonic flow components evolve from the initial geometry of the fluctuating initial conditions  \cite{Alver:2010gr,Alver:2010dn,Bhalerao:2011bp,Bhalerao:2011yg,Teaney:2010vd,Qin:2010pf,Gardim:2011xv} to the final spectrum of observed particles. In other words, the elliptic flow, $v_2$, would be mostly created by the so called participant eccentricity \cite{Alver:2006wh}, $\varepsilon_2$, as well as the triangular flow, $v_3$, would be mostly created by the participant triangularity \cite{Alver:2010gr}, $\varepsilon_3$, and so on. 

It has been observed for Monte Carlo Glauber initial conditions that the Fourier coefficients $v_2$ and $v_3$ show a strong linear dependence with the respective eccentricities $\varepsilon_2$ and $\varepsilon_3$ \cite{Qiu:2011iv,Niemi:2012aj}. However, similar results are not generally observed for higher Fourier coefficients such as, for instance, $v_4$ and $v_5$ \cite{Gardim:2011xv,Teaney:2012ke}. It is important to mention that hydrodynamics, which is widely used to connect the initial conditions to the final spectrum of particles, does not guarantee such a linear response to the initial geometry due to intrinsic nonlinearities present in the hydrodynamic equations.

The aim of this article is to improve the current understanding of the role played by the higher order flow components in the determination of the azimuthal profile of the di-hadron correlation function in heavy ion collisions. The azimuthal component of this function can be parametrized in terms of the Fourier coefficients of the azimuthal distribution of particles, i.e., it can be described in terms of the pair $\left\{v_n,\Psi_n \right\}$ where the first parameter is related to the intensity of the flow component and the second one is an angle that fixes the orientation of the respective harmonic. In particular, we are interested in the situation where the set of triggers is not identical to the set of associated particles, as it occurs when the triggers and the associated particles are defined within different ranges of transverse momentum. In this situation, as we are going to show in Section \ref{sec:introduction_2}, the di-hadron correlation function becomes particularly sensitive to not only the $v_n$ coefficients but the $\Psi_n$ angles as well. More precisely, it depends on the relative phase $\Delta_n =\Psi_n^t-\Psi_n^a$, where the first angle is computed using the triggers and the second using the associated particles. Recently, the fluctuation of this relative phase has been studied in 3+1 ideal hydrodynamics by the NexSPheRIO collaboration \cite{Gardim:2012im} and also in 2+1 viscous hydrodynamics in Ref.\ \cite{Heinz:2013bua,Qiu:2012uy}. In particular, in this article we try to understand the width of the $\Delta_n$ distribution in terms of the geometry of the initial conditions. More precisely, we compute the distribution of the difference $\delta_n =\Psi_n - \Phi_{2,n}$ (in three ranges of $p_T$) in order to quantify the fluctuation of the flow component phase with respect to the geometrical orientation angle $\Phi_{2,n}$, obtained from the initial energy density distribution (in section \ref{sec:introduction_3} the definition of the geometric parameters from initial conditions is discussed). For instance, for Au+Au collisions at $\sqrt{s}_{NN}=200$ GeV, it was shown in Ref. \cite{Gardim:2011re} that the $\delta_n$ distribution, for integrated $p_T$,  is quite narrow (when $n>1$). Here, as we will show, the study of the this distribution for triggers and associated particles can be used to understand the behavior of the $\Delta_n$ distribution. In order to complete the analysis involving flow and initial geometry, we also compute the correlation between the eccentricities $\varepsilon_{2,n}$ and the flow parameters $v_n$ for different bins of transverse momentum.

This paper is organized as follows. In Section \ref{sec:introduction_2} we describe the parametrization of the di-hadron angular correlation function in terms of the Fourier coefficients of the azimuthal distribution of hadrons. In Section \ref{sec:introduction_3} we discuss the definition of the eccentricities used in this paper. In Section \ref{sec:introduction_4} we give the details about our hydrodynamic model including the modeling of the fluctuating initial conditions, the equation of state, and the decoupling mechanism. We discuss our results in Section \ref{sec:results} and we finish in Section \ref{sec:conclusion} with our conclusions. We use a mostly minus metric signature $(+,-,-,-)$ and natural units $\hbar=k_{B}=c=1$.

\section{Fourier Decomposition of the Di-Hadron Angular Correlation Function}\label{sec:introduction_2}

The azimuthal component of the di-hadron correlation function, $C\left(\Delta \phi \right)$, can be defined as follows

\begin{equation}
C\left(\Delta \phi \right)=\frac{\left< \int   g^a  \left(\phi_t + \Delta \phi \right) g^t \left(\phi_t \right) d \phi_t    \right>}{\left< \int g^t \left(\phi_t \right)d \phi_t \right>},
\label{eq:2_part_cf}
\end{equation}

\noindent
where the functions $g^t$ and $g^a$ are the azimuthal distributions of triggers and associated particles in each event, respectively. Each function is associated to an interval of transverse momentum $p_T$. The brackets indicate the average over events (an arithmetic mean). The denominator, naturally, is the average number of triggers.

The decomposition of Eq.\ (\ref{eq:2_part_cf}) in terms of the Fourier coefficients of the azimuthal distribution of hadrons can be obtained using the following expansions

\begin{eqnarray}
g^a \left(\phi_t + \Delta \phi \right) && = v_{0}^a \nonumber\\
&&+\sum_n 2 v_{0}^a v_{n}^a \cos \left[n \left(\phi_t + \Delta \phi - \Psi_{n}^a \right) \right]
\label{eq:fo_a}
\end{eqnarray}

\noindent
and

\begin{equation}
g^t \left(\phi_t \right) = v_{0}^t + \sum_m 2 v_{0}^t v_{m}^t \cos \left[m \left(\phi_t - \Psi_{m}^t \right) \right]\,.
\label{eq:fo_t}
\end{equation}

\noindent
Observe that the values of the parameters $\left\{v_n,\Psi_n \right\}$, for both triggers and associated particles, vary from event to event.

Inserting the expansions (\ref{eq:fo_a}) and (\ref{eq:fo_t}) in Eq.\ (\ref{eq:2_part_cf}), a straightforward calculation leads to the following general formula for the di-hadron correlation function in the azimuthal direction

\begin{equation}
C\left(\Delta \phi \right)=  c_0 + \sum_n c_n \cos(n \Delta \phi) + \sum_n \tilde{c}_n \sin(n \Delta \phi)
\label{eq:2_part_cf_fo}
\end{equation}

\noindent
where

\begin{equation}
c_0 =  \frac{\left<v_{0}^a v_{0}^t\right>}{\left<v_{0}^t\right>},
\label{eq:c0}
\end{equation}

\begin{equation}
c_n = \frac{2}{\left<v_{0}^t\right>}   \left<v_{0}^a v_{0}^t v_{n}^a v_{n}^t   \cos\left[n\left(\Psi_{n}^t - \Psi_{n}^a \right) \right] \right>,
\label{eq:cn}
\end{equation}

\noindent
and

\begin{equation}
\tilde{c}_n = \frac{2}{\left<v_{0}^t\right>} \left<v_{0}^a v_{0}^t v_{n}^a v_{n}^t  \sin\left[n\left(\Psi_{n}^a - \Psi_{n}^t \right) \right]\right>.
\label{eq:cnt}
\end{equation}

Considering the simplest case in which the ranges of transverse momentum for both triggers and associated particles are the same, i.e., the case in which $\Psi_{n}^t = \Psi_{n}^a$, these equations tell us that the profile of the di-hadron correlation function depends only on the $v_{n}$ coefficients, while the odd coefficients, $\tilde{c}_{n}$, are identically null.

On the other hand, when the ranges of transverse momentum for triggers and associated particles are different, $\Psi_{n}^t \ne \Psi_{n}^a$, the following questions can be posed:

\begin{itemize}
\item {What is the profile of the distribution of the relative phase $\left(\Psi_{n}^t - \Psi_{n}^a \right)$ as a function of the transverse momentum and centrality?}

\item {Is the relative phase independent on the $v_n$ coefficients? In other words, can we consider

\begin{eqnarray}
c_n \approx c_n^f = \frac{2}{\left<v_{0}^t\right>} \left<v_{0}^a v_{0}^t v_{n}^a v_{n}^t \right> \hspace{0.5cm} \nonumber\\  \times \left<\cos\left[n\left(\Psi_{n}^t - \Psi_{n}^a \right) \right] \right>
\label{eq:cnf}
\end{eqnarray}

\noindent
when the number of events is sufficiently large (and similarly for the $\tilde{c}_n$ coefficients)? The index $f$ in Eq. (\ref{eq:cnf}) indicates the factorization of the $c_n$ coefficient. Note that even when this factorization is valid, the average over the cosine still needs to be determined.}
\end{itemize}

It must be mentioned that in the well-known event plane method \cite{Poskanzer:1998yz} $\Psi_n^t = \Psi_n^a = \Psi_n^{\mathbf{EP}}$ regardless of the $p_T$ bin chosen for the trigger and associated hadrons, which means that the relative phase in this case is identically zero and, consequently, the di-hadron correlation function is necessarily an even function of $\Delta\phi$. However, in the case where the triggers and associated particles are defined within different $p_T$ bins, there is no reason to assume that $\Psi_n^t$ and $\Psi_n^a$ are aligned in every single event. On the contrary, it is more natural to suppose that the relative phase fluctuates from event to event. In this scenario, considering that the relative phase distribution shows a peak at the origin $\Delta_n=0$ with some width, the main question becomes how far from the unit the absolute value of the factor $\cos \left[n \left(\Psi_n^t-\Psi_n^a \right) \right]$ is. As one can see in Eq. (\ref{eq:cn}), this factor can change the di-hadron angular correlation profile.

When the number of events is sufficiently large one expects that factorization can be used in Eqs. (\ref{eq:cn}) and (\ref{eq:cnt}), and the resulting factor $\left\langle \sin \left[n \left(\Psi_n^a-\Psi_n^t \right) \right] \right\rangle$, related with the odd coefficient, is expected to average to zero. This means that the relative phase distribution becomes an even function in this limit and, thus, the parity of the di-hadron correlation function $C\left(\Delta \phi \right)$ is restored.

\section{Eccentricity definition}\label{sec:introduction_3}

In order to quantify the anisotropy of the initial conditions event by event, in this article we will use the following definition for the eccentricities

\begin{equation}
\varepsilon_{m,n} = \frac{\left\{r^m \cos \left[n \left(\phi - \Phi_{m,n} \right)  \right]  \right\}}{\left\{r^m \right\}}\,,
\label{eq:eccentricity}
\end{equation}

\noindent
where $\{...\}$ indicates the average weighted by the energy density profile $\epsilon \left(\vec{r} \right)$ (see Fig.\ \ref{fig:ic1}) in the transverse plane. The corresponding orientation angle is given by

\begin{equation}
\Phi_{m,n}=\frac{1}{n} \tan^{-1} \left( \frac{\left\{r^m \sin \left(n \phi \right) \right\}}{\left\{r^m \cos \left(n \phi \right) \right\}} \right)\,.
\end{equation}

\noindent
Finally, $r^m = \left(x^2+y^2\right)^\frac{m}{2}$ and $\phi=\tan^{-1} \left(y/x \right)$.

The index $m$ can be conveniently chosen to improve the prediction of the respective flow component. For instance, in Ref. \cite{Gardim:2011xv}, in the context of the NexSPheRIO code \cite{Hama:2004rr}, it was shown that the triangular flow, $v_3$, is better predicted using $m=3$. However, in this article we follow the original proposal in \cite{Alver:2010gr} and fix $m=2$. It is easy to verify that $\varepsilon_{2,2}$ is the well-known participant eccentricity \cite{Alver:2006wh} in a coordinate system where $\left<x \right>=\left<y\right>=0$. Thus,

\begin{equation}
\varepsilon_{2,2}=\frac{\sqrt{\left(\sigma_x^2 - \sigma_y^2 \right)^2+4 \sigma_{xy}^2}}{\sigma_x^2 + \sigma_y^2},
\end{equation}

\noindent
where $\sigma_x^2 = \left<x \right>^2$, $\sigma_y^2 = \left<y \right>^2$ and $\sigma_{xy}^2 = \left<xy \right>^2$.\\

A motivation for the definition (\ref{eq:eccentricity}) can be found in Ref. \cite{Teaney:2010vd} where it was shown that the eccentricities defined above are related to the irreducible components of the cumulant expansion of the initial energy density distribution.

Given the eccentricities and their respective orientation angles, some interesting questions can be posed:

\begin{itemize}

\item{Do the Fourier coefficients, $v_n$, show a linear dependence on the respective eccentricity, $\varepsilon_{2,n}$, independently on the transverse momentum range and centrality?}

\item{Are the angles $\Phi_{2,n}$ and $\Psi_n$ aligned?}

\end{itemize}

As it was mentioned in the previous section, Eqs.\ (\ref{eq:cn}) and (\ref{eq:cnt}) show that the $v_n$ coefficients alone (or, equivalently, the eccentricities alone) do not provide enough information to produce the azimuthal structures observed in the di-hadron correlation function \cite{Adare:2008ae} in the case where the $p_T$ bins of the triggers and associated hadrons are different. For instance, in an extreme case in which the relative phases are randomly distributed, the di-hadron correlation function would not show any structure independently on the value of the $v_n$ coefficients. In this context, a partial alignment between the orientation angle $\Phi_{2,n}$ and the phase $\Psi_n^a$ (see, for instance, Ref. \cite{Gardim:2011re}), as well as a partial alignment between $\Phi_{2,n}$ and $\Psi_n^t$, would indicate a partial alignment between the angles $\Psi_n^a$ and $\Psi_n^t$.

\section{\label{sec:introduction_4} Details of the Hydrodynamic Modelling}

In this work, we use a (2+1) (i.e., boost-invariant \cite{Bjorken:1982qr}) ideal relativistic fluid to study the connection between the initial conditions and final flow observables relevant to the description of the di-hadron angular correlation function. We are focusing on the transverse expansion near mid-rapidity. In practice, we consider only a thin transverse slice of matter, so that $|y|<0.12$, where $y$ is the rapidity. In order to solve the ideal hydrodynamic equations, we apply the relativistic version of the so called Smoothed Particles Hydrodynamics (SPH) approach originally developed in \cite{Aguiar:2000hw}, which is a suitable tool to deal with irregular distributions of matter (details about the SPH method and a discussion of how the ideal fluid nonlinear partial differential equations are solved within this approach are given in Appendix \ref{sec:sph}). We assume that the baryon chemical potential is zero. Moreover, the initial transverse velocity is set zero. Our code matches the previous tests made using the NexSPheRIO code \cite{Hama:2004rr} and, in Appendix \ref{sec:gubser}, we show that our code is able to perfectly match the exact solution of 2+1 ideal hydrodynamics obtained in Ref. \cite{Gubser:2010ze} (also known as Gubser flow).

\begin{figure}
\includegraphics[scale=0.50]{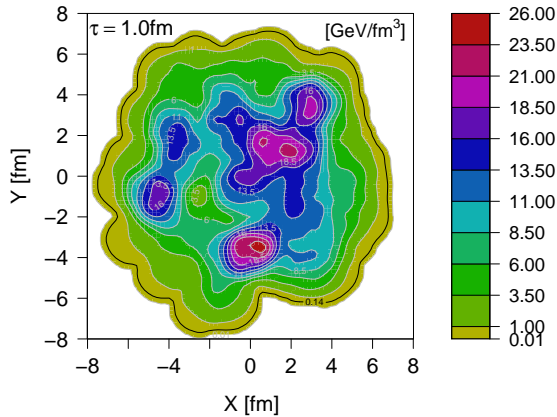}% Here is how to import EPS art
\caption{\label{fig:ic1}Initial energy density distribution in the transverse plane at the mid-rapidity for a randomly chosen Au+Au collision at 200GeV in the $(0-5)\%$ centrality class, computed using an implementation of the Monte Carlo Glauber Model developed in \cite{Drescher:2006ca,Drescher:2007ax}.}
\end{figure}

In order to get an idea of the type of energy density profiles obtained in event by event simulations, we show in Fig. \ref{fig:ic1} the initial energy density distribution in the transverse plane at the mid-rapidity for a randomly chosen central Au+Au collision at 200GeV, computed using an implementation of the Monte Carlo Glauber model developed in \cite{Drescher:2006ca,Drescher:2007ax} and used throughout this work. Note that this distribution is quite irregular showing several regions where the energy is considerably concentrated (the so called hot spots). Since the initial anisotropy in this model arises basely from the random position of the incident nucleons, the regions where the energy is concentrated correspond to the regions where the nucleon density is large. There is a normalization factor associated to the initial energy density distribution, which is chosen through a comparison to data. We set this factor so that the maximum of the average temperature distribution, in the $(0-5)\%$ centrality window, coincides with the temperature of 0.31 GeV (similar values can be found in the literature; see, for instance, Ref. \cite{Luzum:2010ae}). Once fixed by the central collisions, this overall factor is kept the same for the peripheral collisions studied in this work.

We use the equation of state EOS S95n-v1 \cite{Huovinen:2009yb} in our model, which combines results from lattice QCD at high temperatures and the hadron resonance gas equation at low temperatures. The decoupling mechanism is based on the Cooper-Frye prescription \cite{Cooper:1974mv}. In this approach, the particles become free after crossing a hyper-surface of constant temperature, denominated freeze-out temperature, $T_{fo}$ (the details about the Cooper-Frye prescription in the SPH approach are discussed in Appendix \ref{sec:cfpsph}). In our hydrodynamic code, we have not implemented the decay of particles yet.  All the results presented in this article correspond then to direct positively charged pions.

Since the goal of this article is not to make a rigorous comparison between our numerical results and the data but rather to understand how the flow components create the structures observed in the di-hadron correlation function, the role of the freeze-out temperature here is just to determine the expansion time of the fluid. By using $T_{fo}=0.14$GeV, which is a typical value in the literature (see, for instance, Ref. \cite{Luzum:2010ae}), the total expansion time in the $\left(0-5 \right)\%$ centrality window is around 15 fm. Proportionally,  in the $\left(20-30 \right)\%$, the expansion time is shorter $\sim 10$ fm. In both centrality windows, studied in this article, the expansion time is sufficiently long to induce the hydrodynamic effects in the final spectrum of hadrons.

The initial time at which we begin the hydrodynamic evolution is $\tau_0=1$ fm. In this work, the smoothing SPH parameter is chosen to be $h=0.3$ fm (see the discussion in Appendix \ref{sec:sph}), which allows for relatively quick computation times while still preserving the important structure present in the initial conditions. 

Summarizing, the procedure to compute an observable in a single event is the following: (i) Monte Carlo Glauber initial conditions are used to obtain the initial energy density in the transverse plane; (ii) the hydrodynamic evolution is calculated using the SPH method \cite{Aguiar:2000hw} and (iii) the final spectra is computed using the Cooper-Frye prescription \cite{Cooper:1974mv}. At the end of the simulation, the average value of a given observable is calculated over an ensemble of events. All the results presented in this article correspond to Au+Au collisions at $\sqrt{s}_{NN}$= 200 GeV and 1000 events were computed.

\section{\label{sec:results} Results}

\begin{figure*}
\includegraphics[scale=0.60]{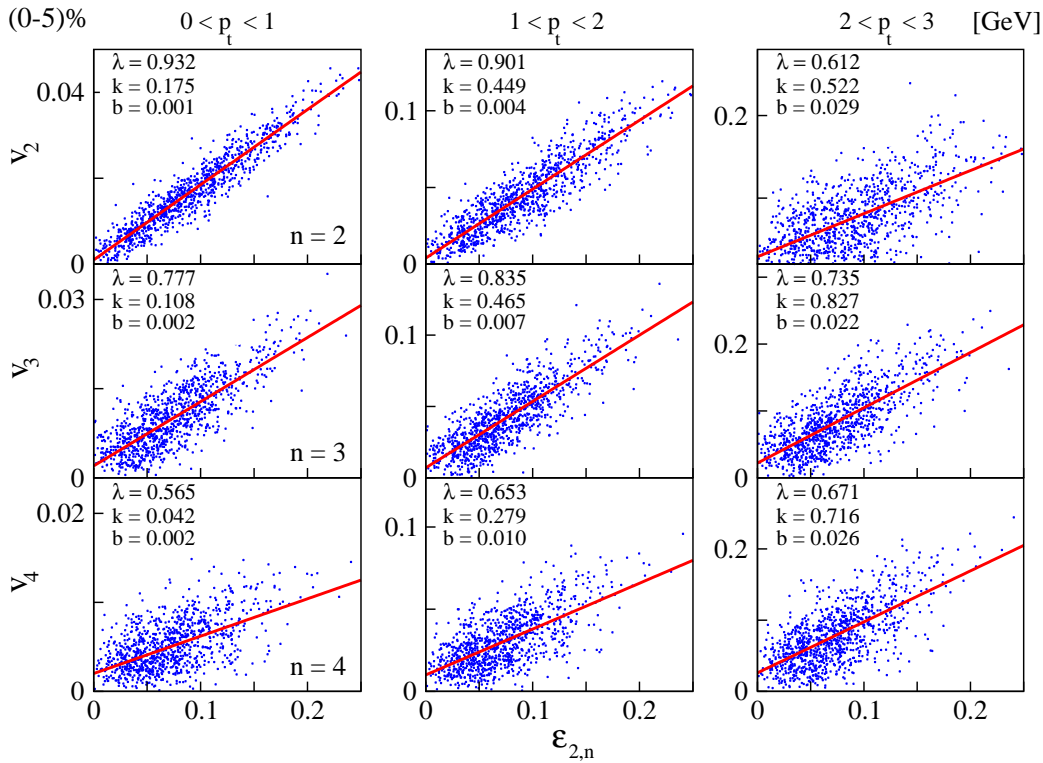}% Here is how to import EPS art
\caption{\label{fig:vn_x_emn_0_5} Correlation between the initial eccentricity, $\varepsilon_{2,n}$  (n=2,3,4), and the respective harmonic coefficient, $v_n$, in the (0-5)$\%$ centrality window with 1000 events. Three ranges of transverse momentum are presented. The parameters $k_n$ and $b_n$ are obtained from the linear fit: $v_n = k_n \varepsilon_{2,n} + b_n$. The parameter $\lambda_n$ is the linear correlation coefficient, see Eq. (\ref{equ:lcoeff}).}
\end{figure*}

\begin{figure*}
\includegraphics[scale=0.60]{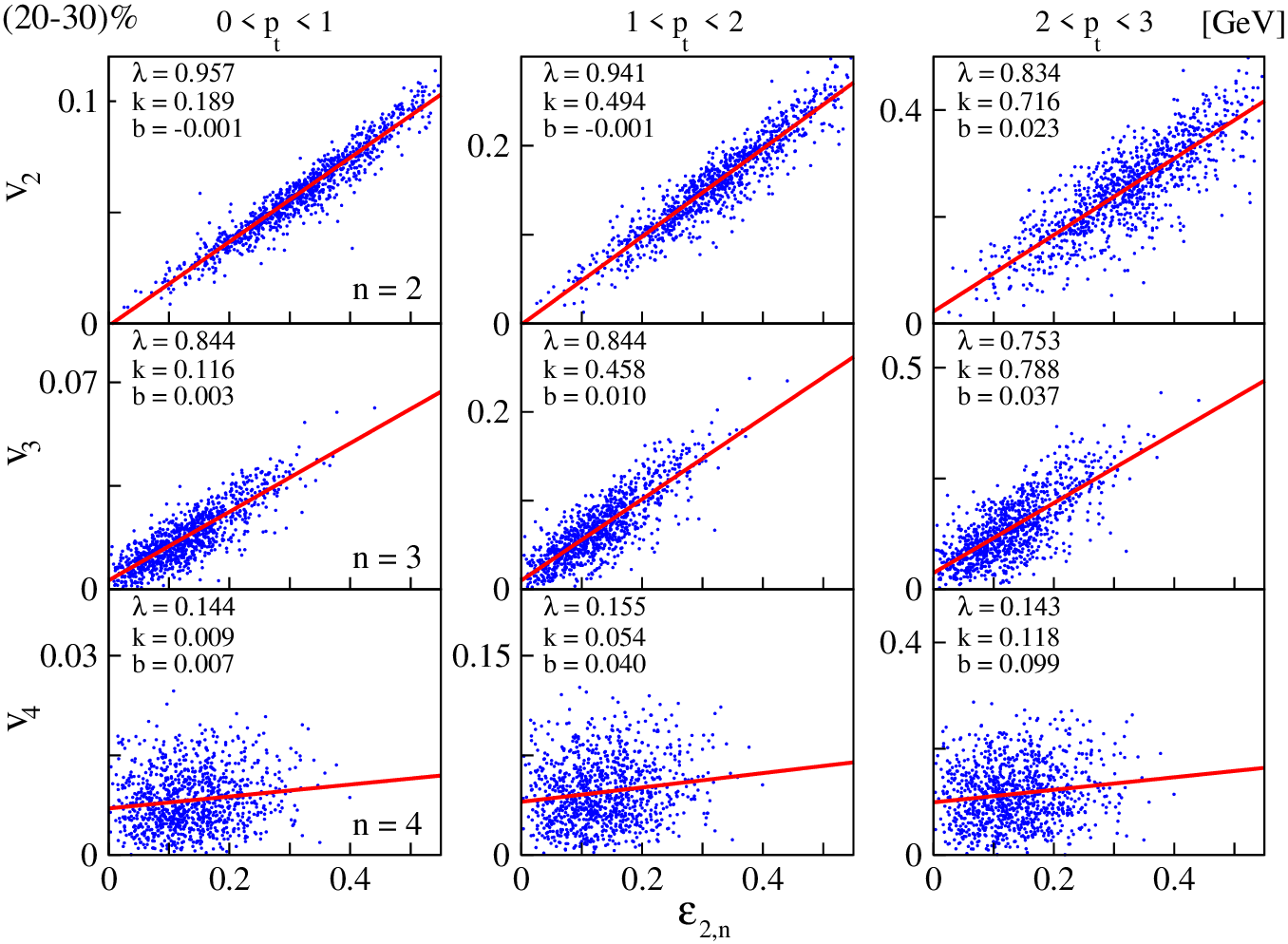}% Here is how to import EPS art
\caption{\label{fig:vn_x_emn_20_30} Correlation between the initial eccentricity, $\varepsilon_{2,n}$  (n=2,3,4), and the respective harmonic coefficient, $v_n$, in the (20-30)$\%$ centrality window with 1000 events. Three ranges of transverse momentum are presented. The parameters $k_n$ and $b_n$ are obtained from the linear fit: $v_n = k_n \varepsilon_{2,n} + b_n$. The parameter $\lambda_n$ is the linear correlation coefficient, see Eq. (\ref{equ:lcoeff}).}
\end{figure*}

In Fig.\ \ref{fig:vn_x_emn_0_5} we show the correlation between the initial eccentricity, $\varepsilon_{2,n}$  $(n=2,3,4)$, and the respective flow coefficient, $v_n$, in the (0-5)$\%$ centrality window (central). Three ranges of transverse momentum are presented. A similar graph is shown in Fig.\ \ref{fig:vn_x_emn_20_30}, in the (20-30)$\%$ centrality window (peripheral). The parameters $k_n$ and $b_n$ are obtained from the linear fit: $v_n = k_n \varepsilon_{2,n} + b_n$. Moreover,

\begin{equation}
\lambda_n=\frac{\left< \left(\varepsilon_{2,n} -  \left< \varepsilon_{2,n} \right> \right) \left(v_n -  \left< v_n \right> \right)  \right> }{ \sqrt{\left< \left(\varepsilon_{2,n} -  \left< \varepsilon_{2,n} \right> \right)^2 \right> \left< \left(v_n -  \left< v_n \right> \right)^2 \right> }}
\label{equ:lcoeff}
\end{equation}

\noindent
is the linear correlation coefficient. The closer to the unit $\left| \lambda_n \right|$ is, the stronger the linear correlation between the variables $\varepsilon_{2,n}$ and $v_n$ becomes. In fact, when $\lambda\sim 1$  ($\lambda\sim -1$) both variables show a strong linear correlation (anti-correlation).

One can see that the coefficients $v_2$ and $v_3$ are considerably correlated (linear correlation) with respect to the eccentricities $\varepsilon_{2,2}$ and $\varepsilon_{2,3}$, respectively. This behavior is observed in almost all cases - for high transverse momentum particles ($2 < p_t< 3$ GeV), in the $\left(0-5 \right)\%$ centrality window, the parameter $\lambda_2$ is smaller in comparison to the other cases ($\lambda_2=0.612$). In particular, $\lambda_2$ obtained in the $\left(20-30 \right)\%$ centrality window is closer to the unit in comparison to the same parameter obtained in the $\left(0-5 \right)\%$ centrality window, due to the almond-like transverse shape of the initial conditions in the peripheral window, which produces stronger elliptic flow.

On the other hand, $\lambda_3$ is less sensitive to centrality, which supports the idea that $\varepsilon_{2,3}$ is driven by fluctuations. This shows that the almond shape of the initial conditions in the $\left(20-30 \right)\%$ centrality window does not interfere with the correlation between $\varepsilon_{2,3}$ and $v_3$. Finally, the linear correlation between $\varepsilon_{2,4}$ and $v_4$ is weaker, especially for peripheral collisions. These results are compatible with those obtained in Ref. \cite{Niemi:2012aj} where the linear correlation between the $p_T$ integrated flow coefficients and the eccentricities was investigated (within viscous hydrodynamics).

\begin{figure*}
\includegraphics[scale=0.54]{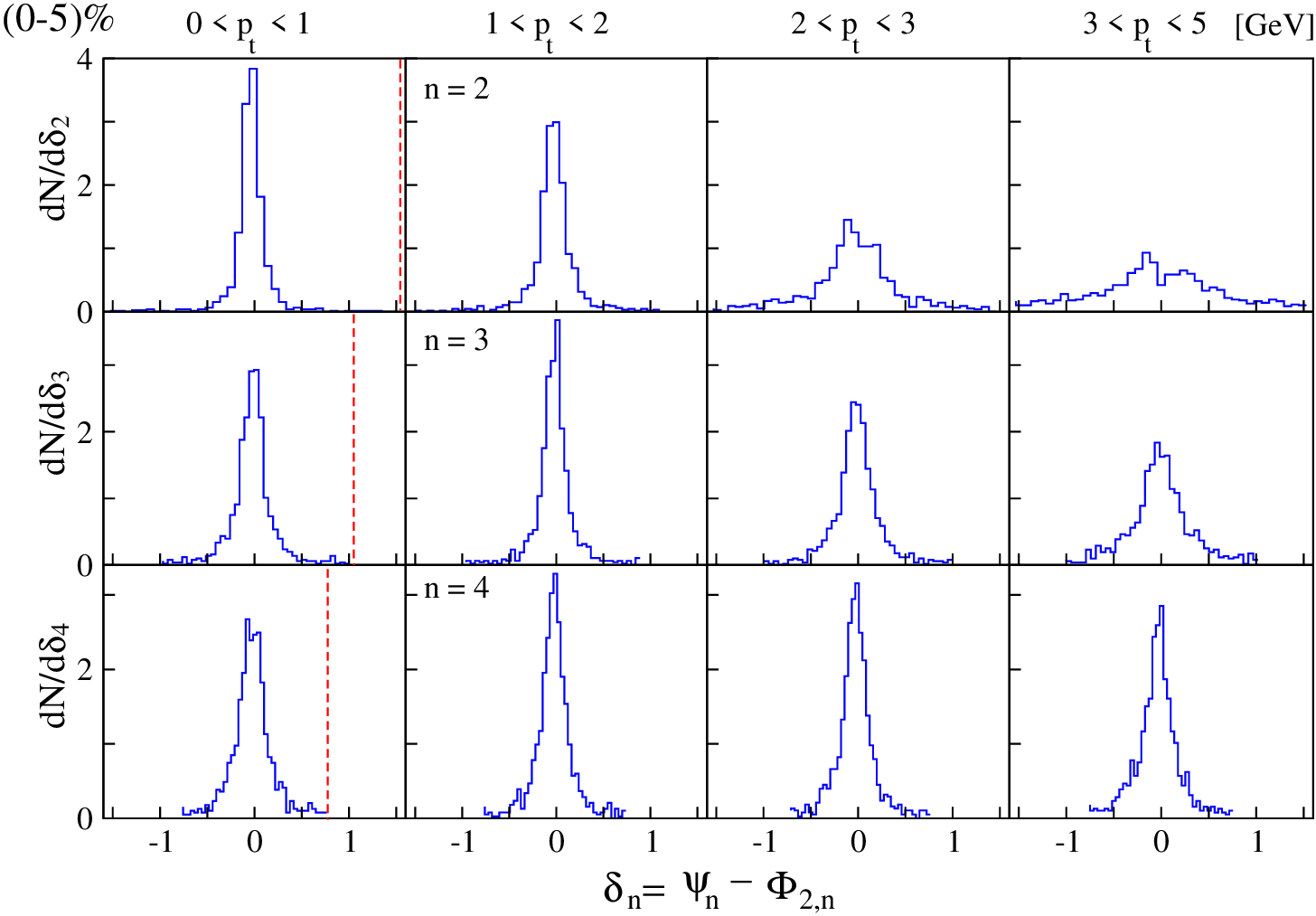}% Here is how to import EPS art
\caption{\label{fig:dN_ddeln_x_deln_0_5} Distribution of the angular difference $\delta_n = \Psi_n - \Phi_{2,n}$ $(n=2,3,4)$ in the (0-5)$\%$ centrality window with 1000 events. Four ranges of transverse momentum are presented. All distributions are normalized to one. The vertical dashed  line indicates the maximal difference for each harmonic, i.e., $\delta_n^{\mathbf{max}} =\pi/n$.}
\end{figure*}

\begin{figure*}
\includegraphics[scale=0.54]{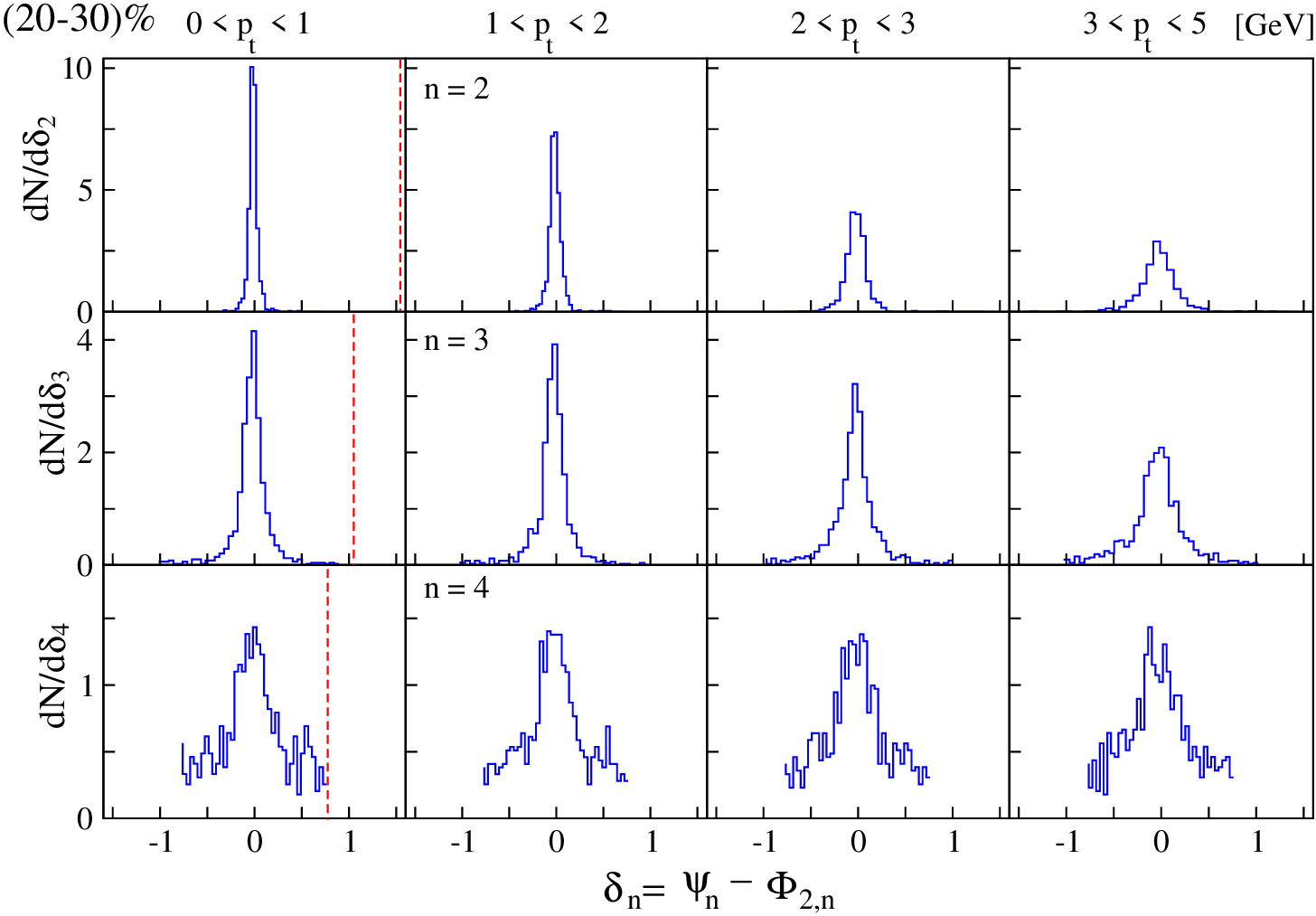}% Here is how to import EPS art
\caption{\label{fig:dN_ddeln_x_deln_20_30} Distribution of the angular difference $\delta_n=\Psi_n - \Phi_{2,n}$ in the (20-30)$\%$ centrality window with 1000 events. Four ranges of transverse momentum are presented. All distributions are normalized to one. The vertical dashed  line indicates the maximal difference for each harmonic, i.e., $\delta_n^{\mathbf{max}} =\pi/n$.}
\end{figure*}

In Fig.\ \ref{fig:dN_ddeln_x_deln_0_5} we show the distribution of the angular difference $\delta_n = \Psi_n - \Phi_{2,n}$ $(n=2,3,4)$ in the (0-5)$\%$ centrality window. Note that $\Psi_n$ is rotated by $\pi/n$ in order to achieve the smallest angular difference with respect to $\Phi_n$. Four ranges of transverse momentum are presented. A similar graph is shown in Fig. \ref{fig:dN_ddeln_x_deln_20_30} for the (20-30)$\%$ centrality window. All distributions are normalized to one. The vertical dashed  line indicates the maximal difference for each harmonic, i.e., $\delta_n^{\mathbf{max}} =\pi/n$. These results show that there is a partial alignment between the initial reference angle $\Phi_{2,n}$ and the flow angle $\Psi_n$ in almost all of cases. For high transverse momentum particles ($2 < p_t< 5$ GeV), in the $\left(0-5 \right)\%$ centrality window, the difference $\delta_2$ is broader in comparison to the remaining cases.

Note that the almond-like transverse shape of the initial conditions in the $\left(20-30 \right)\%$ centrality window produces a stronger elliptic flow that reduces the fluctuation of the angle $\Psi_2$ with respect to the reference angle $\Phi_{2,2}$, making the distribution of the difference $\delta_2$ narrower. On the other hand, this mechanism does not influence the partial alignment between $\Psi_3$ and $\Phi_{2,3}$, which is quantified by the width of the distribution of the difference $\delta_3$, i.e., in both centrality windows $\delta_3$ is narrow. With respect to the forth harmonic, $\Psi_4$ and $\Phi_{2,4}$ show a partial alignment as well (event though the linear correlation between $v_4$ and $\varepsilon_{2,4}$ is weak).

\begin{figure*}
\includegraphics[scale=0.54]{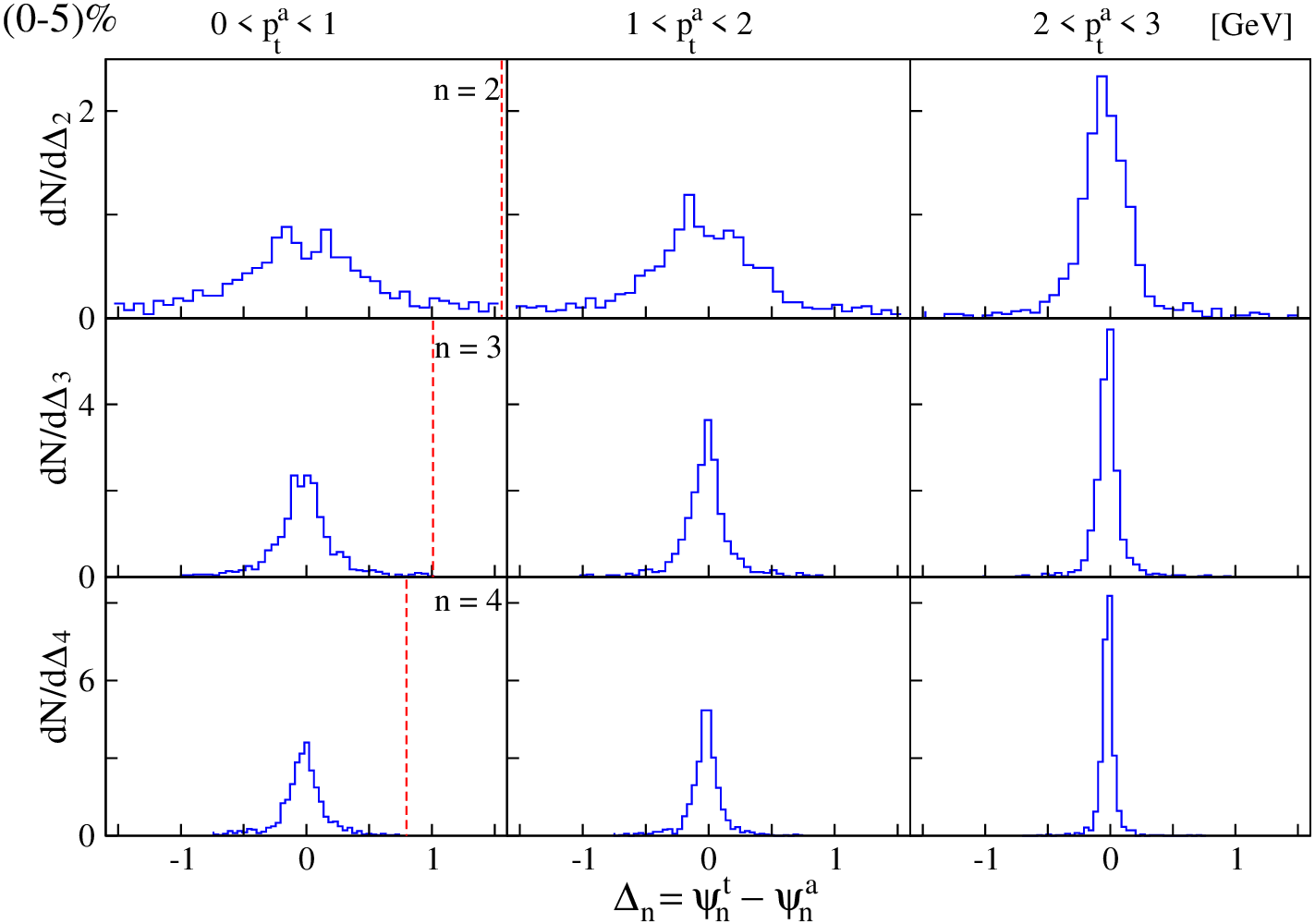}% Here is how to import EPS art
\caption{\label{fig:dN_dDn_x_Dn_0_5} Distribution of the phase difference $\Delta_n = \Psi_n^t - \Psi_n^a$ (n=2,3,4) in the (0-5)$\%$ centrality window after 1000 events. Three ranges of transverse momentum, for associated particles, are presented. All distributions are normalized to one. The vertical dashed  line indicates the maximal difference for each harmonic, i.e., $\Delta_n^{\mathbf{max}} =\pi/n$. The range in transverse momentum for the triggers is defined as $3 < p_t^{\mathbf{trigg}}< 5$ GeV.}
\end{figure*}

\begin{figure*}
\includegraphics[scale=0.54]{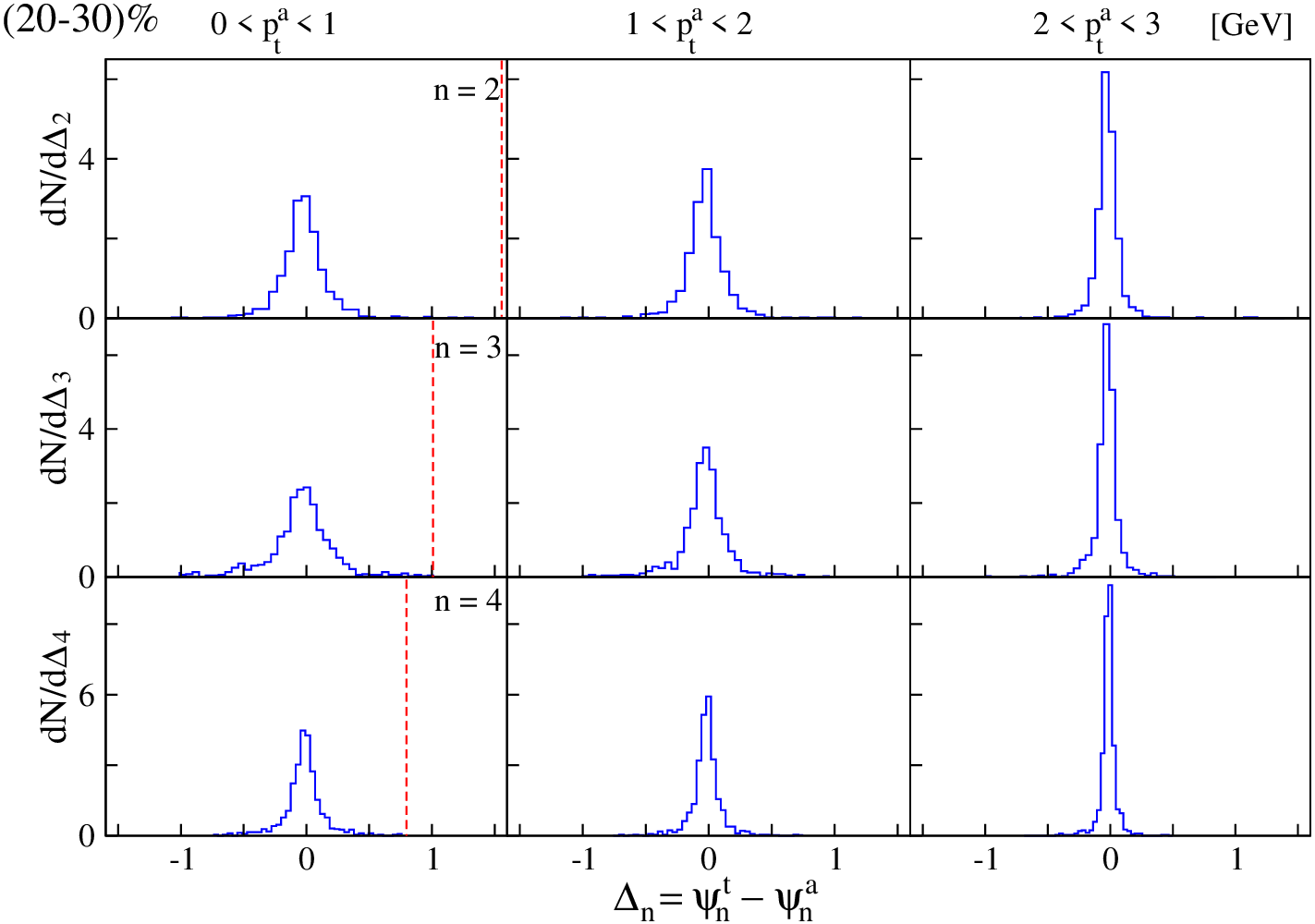}% Here is how to import EPS art
\caption{\label{fig:dN_dDn_x_Dn_20_30} Distribution of the phase difference $\Delta_n = \Psi_n^t - \Psi_n^a$ (n=2,3,4) in the (20-30)$\%$ centrality window after 100 events. Three ranges of transverse momentum, for associated particles, are presented. All distributions are normalized to one. The vertical dashed  line indicates the maximal difference for each harmonic, i.e., $\Delta_n^{\mathbf{max}} =\pi/n$. The range in transverse momentum for the triggers is defined as $3 < p_t^{\mathbf{trigg}}< 5$ GeV.}
\end{figure*}

Now we come to study of the phase difference between triggers and associated hadrons. In Fig.\ \ref{fig:dN_dDn_x_Dn_0_5} the distribution of the relative phase $\Delta_n = \Psi_n^t - \Psi_n^a$ (n=2,3,4) is shown for the (0-5)$\%$ centrality window. Three ranges of transverse momentum for associated particles are presented. A similar graph is shown in Fig.\ \ref{fig:dN_dDn_x_Dn_20_30} for the (20-30)$\%$ centrality window. All distributions are normalized to one. The vertical dashed  line indicates the maximal difference for each harmonic, i.e., $\Delta_n^{\mathbf{max}} =\pi/n$. The range of transverse momentum for the triggers is defined as $3 < p_t^{\mathbf{trigg}}< 5$ GeV. In Ref.\ \cite{Mota:2012qv} a similar observable was employed to investigate the granularity of the initial conditions. 

These results show that there is also a partial alignment between the angles $\Psi_n^t$ and $\Psi_n^a$. Starting from the left and going to right side in these plots, the distribution of the difference $\Delta_n$ tends to be narrower once the kinematic region of the associated particles gets close to the kinematic region of the triggers. However, when both kinematic regions are far from each other, for instance by choosing the associated particles with low transverse momentum ($0 < p_t^a< 1$ GeV), this distribution can become considerably broad. This is the case for the $\Delta_2$ distribution computed in the $\left(0-5 \right)\%$ centrality window. As we shall show below, this behavior makes the factor $\cos \left[2 \left(\Psi_2^t - \Psi_2^a  \right) \right]$ deviates significantly from the unit. In contrast, the same distribution is narrow in the $\left(20-30 \right)\%$ centrality window. This behavior can be understood in terms of the geometry of the initial conditions. For instance, in Fig.\ \ref{fig:dN_ddeln_x_deln_0_5} one can see that the difference $\delta_2$ for the triggers is broad while the same distribution for the associated particles is narrow, which means that $\Psi_2^a$ and $\Psi_2^t$ are not always aligned. For the remaining cases (with $0 < p_t^a< 1$ GeV), the relative phase distribution is narrower because both angles are better aligned with the reference angle.

We show in Table \ref{tab:table2} the average of the factors $\left< \cos \left(n \Delta_n  \right) \right>$ and $\left< \sin \left(n \Delta_n  \right) \right>$ for the first seven harmonics, within three ranges of transverse momentum of the associated particles, in the $\left(0-5 \right)\%$ centrality window. The Table \ref{tab:table3} shows the same quantities for the $\left(20-30 \right)\%$ centrality window. The values in both tables are related to the width of the distributions that are shown in Figs.\ \ref{fig:dN_dDn_x_Dn_0_5} and \ref{fig:dN_dDn_x_Dn_20_30}, respectively.

As one can see from the tables, the sine factors average to zero, as expected. This means that both the relative phase distribution and the di-hadron correlation function are even functions. With respect to the cosine factors, the absolute values are smaller than the unit as a consequence of the fluctuations. The remarkable case occurs for associated particles with low transverse momentum ($0 < p_t^a< 1$ GeV) in the $\left(0-5 \right)\%$ centrality window where $\left< \cos \left(2 \Delta_2  \right) \right> = 0.456$. The negative signal that appears associated to the first harmonic is a consequence of the conservation of the momentum in the transverse plane -  if the associated particles move in one direction, the triggers must move in the opposite direction to conserve momentum.

\begin{table}[t]
\caption{\label{tab:table2} Average of the factors $\cos \left(n \Delta_n  \right)$ and $\sin \left(n \Delta_n  \right)$ for three ranges of transverse momentum of the associated particles in the $\left(0-5 \right)\%$ centrality window after 1000 events. The range in transverse momentum for the triggers is defined as $3 < p_t^{\mathbf{trigg}}< 5 GeV$. The first seven harmonics are shown.}
\begin{ruledtabular}
\begin{tabular}{lrrr}
$\left < \cos (n \Delta_n) \right>$ & $0 < p_t^a < 1$ & $1 < p_t^a < 2$ & $2 < p_t^a < 3$ \\
\hline
$n=1$ & -0.741 & 0.809 & 0.969 \\
$n=2$ & 0.456 & 0.578 & 0.848 \\
$n=3$ & 0.766 & 0.842 & 0.932 \\
$n=4$ & 0.766 & 0.857 & 0.955 \\
$n=5$ & 0.759 & 0.836 & 0.956 \\
$n=6$ & 0.811 & 0.861 & 0.952 \\
$n=7$ & 0.842 & 0.875 & 0.957 \\
\hline
$\left < \sin (n \Delta_n) \right>$ & $0 < p_t^a < 1$ & $1 < p_t^a < 2$ & $2 < p_t^a < 3$ \\
\hline
$n=1$ & -0.020 & 0.002 & 0.005 \\
$n=2$ & -0.004 & 0.011 & 0.004 \\
$n=3$ & -0.004 & -0.007 & -0.009 \\
$n=4$ & 0.012 & 0.020 & 0.012 \\
$n=5$ & 0.010 & 0.012 & 0.005 \\
$n=6$ & 0.004 & 0.001 & 0.003 \\
$n=7$ & -0.016 & -0.009 & -0.001 \\
\end{tabular}
\end{ruledtabular}
\end{table}

\begin{table}[t]
\caption{\label{tab:table3} Average of the factors $\cos \left(n \Delta_n  \right)$ and $\sin \left(n \Delta_n  \right)$ for three ranges of transverse momentum of the associated particles in the $\left(20-30 \right)\%$ centrality window after 1000 events. The range in transverse momentum for the triggers is defined as $3 < p_t^{\mathbf{trigg}}< 5 GeV$. The first seven harmonics are shown.}
\begin{ruledtabular}
\begin{tabular}{lrrr}
$\left < \cos (n \Delta_n) \right>$ & $0 < p_t^a < 1$ & $1 < p_t^a < 2$ & $2 < p_t^a < 3$ \\
\hline
$n=1$ & -0.721 & 0.709 & 0.940 \\
$n=2$ & 0.931 & 0.946 & 0.977 \\
$n=3$ & 0.747 & 0.850 & 0.954 \\
$n=4$ & 0.823 & 0.889 & 0.964 \\
$n=5$ & 0.860 & 0.906 & 0.971 \\
$n=6$ & 0.877 & 0.907 & 0.969 \\
$n=7$ & 0.899 & 0.921 & 0.967 \\
\hline
$\left < \sin (n \Delta_n) \right>$ & $0 < p_t^a < 1$ & $1 < p_t^a < 2$ & $2 < p_t^a < 3$ \\
\hline
$n=1$ & -0.031 & 0.027 & 0.013 \\
$n=2$ & -0.011 & -0.008 & -0.006 \\
$n=3$ & 0.022 & 0.023 & 0.013 \\
$n=4$ & -0.003 & 0.000 & 0.005 \\
$n=5$ & -0.014 & -0.010 & -0.006 \\
$n=6$ & -0.013 & -0.011 & -0.004 \\
$n=7$ & 0.005 & 0.003 & 0.002 \\
\end{tabular}
\end{ruledtabular}
\end{table}

\begin{figure*}
\includegraphics[scale=0.55]{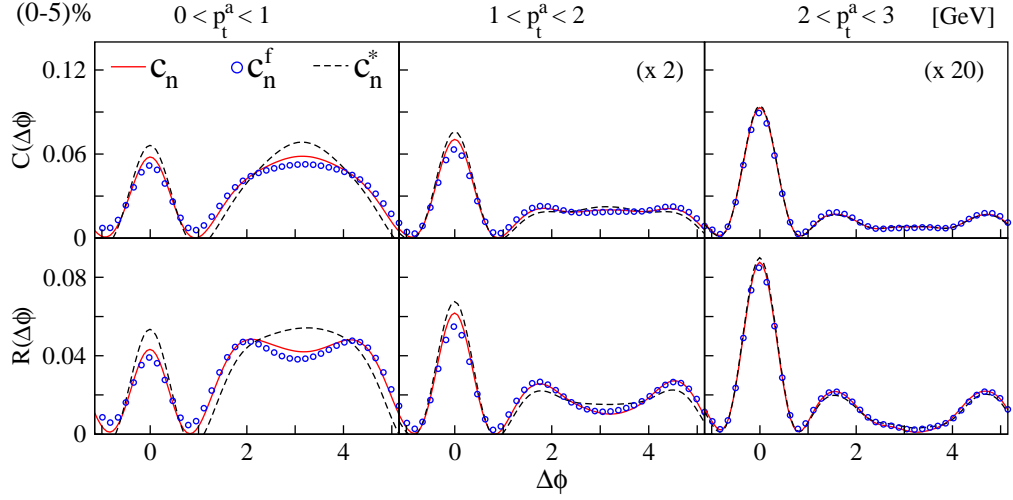}% Here is how to import EPS art
\caption{\label{fig:dN_ddphi_x_dphi_0_5} Total di-hadron correlation function $C \left(\Delta \phi \right)$ (upper panels) and the corresponding background subtracted correlation function $R \left(\Delta \phi \right)$ (lower panels) for three ranges of transverse momentum of the associated particles in the $\left(0-5 \right)\%$ centrality window after 1000 events. The range in transverse momentum for the triggers is defined as $3 < p_t^{\mathbf{trigg}}< 5$ GeV. The solid lines correspond to the formulas in Eqs.\ (\ref{eq:cn}) and (\ref{eq:cnt}) ($c_n$), the lines with circles correspond to the factorized formula (\ref{eq:cnf}) ($c_n^f$) - and analogously for the sine terms - and the dashed lines correspond to the formulas (\ref{eq:cn}) and (\ref{eq:cnt}) without the cosine and sine factors, respectively, ($c_n^*$).}
\end{figure*}

\begin{figure*}
\includegraphics[scale=0.55]{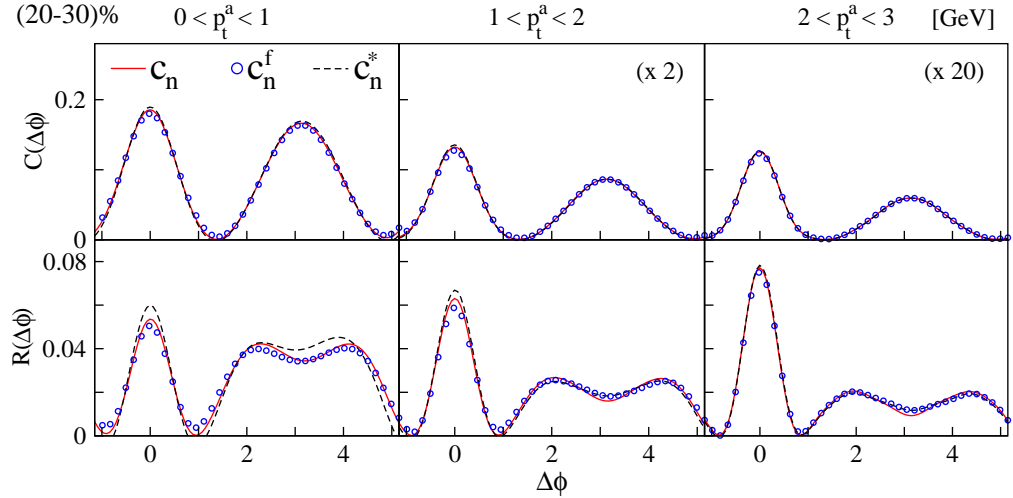}% Here is how to import EPS art
\caption{\label{fig:dN_ddphi_x_dphi_20_30} Total di-hadron correlation function $C \left(\Delta \phi \right)$ (upper panels) and the corresponding background subtracted correlation function $R \left(\Delta \phi \right)$ (lower panels) for three ranges of transverse momentum of the associated particles in the $\left(20-30 \right)\%$ centrality window after 1000 events. The range in transverse momentum for the triggers is defined as $3 < p_t^{\mathbf{trigg}}< 5$ GeV. The solid lines correspond to the formulas in Eqs.\ (\ref{eq:cn}) and (\ref{eq:cnt}) ($c_n$), the lines with circles correspond to the factorized formula (\ref{eq:cnf}) ($c_n^f$) - and analogously for the sine terms - and the dashed lines correspond to the formulas (\ref{eq:cn}) and (\ref{eq:cnt}) without the cosine and sine factors, respectively, ($c_n^*$).}
\end{figure*}

In Fig.\ \ref{fig:dN_ddphi_x_dphi_0_5} we show the total di-hadron correlation function $C \left(\Delta \phi \right)$ and the corresponding background subtracted function $R \left(\Delta \phi \right)$ for three ranges of transverse momentum of the associated particles. The range in transverse momentum for the triggers is kept the same as before. A similar plot is shown in Fig.\ \ref{fig:dN_ddphi_x_dphi_20_30} for the $\left(20-30 \right)\%$ centrality window. The solid lines correspond to the formulas in Eqs.\ (\ref{eq:cn}) and (\ref{eq:cnt}) ($c_n$), the lines with circles correspond to the factorized formula (\ref{eq:cnf}) ($c_n^f$) - and analogously for the sine terms - and the dashed lines correspond to the formulas (\ref{eq:cn}) and (\ref{eq:cnt}) without the cosine and sine factors, respectively, ($c_n^*$).

The method that we used to remove the background and define the function $R \left(\Delta \phi \right)$ is a variation of the widely known mixed event method (this was also used in \cite{Takahashi:2009na}). In this method, the associated particles and the triggers are chosen in different events, producing a mixed correlation. This is usually used to remove the longitudinal correlation that arises from the shape of the longitudinal distribution of particles. In our version of this method, the events, that will be mixed, are aligned according to the direction of the event plane $\Psi_2^{\mathbf{EP}}$. This procedure creates a background of the form $c_2^{\mathbf{mix}} \cos \left(2 \Delta \phi \right)$.

One can see see from Figs.\ \ref{fig:dN_ddphi_x_dphi_0_5} and \ref{fig:dN_ddphi_x_dphi_20_30} that the fluctuation of the relative phases can change the shape of the di-hadron correlation function and the effect becomes more significant when the associated particles are chosen with low transverse momentum in comparison to the triggers. Observe that the factorized formula is already a reasonable approximation to the original formulas, Eqs.\ (\ref{eq:cn}) and (\ref{eq:cnt}), after 1000 events.

\section{\label{sec:conclusion} Conclusions}

In this paper, we studied the di-hadron angular correlation function within boost invariant, ideal hydrodynamics for Au+Au collisions at $\sqrt{s}_{NN}=200$ GeV using Monte Carlo Glauber fluctuating initial conditions. We observed, when $0<p_T< 3$ GeV, that the intensity of the flow components and their phases, $\left\{v_n, \Psi_n \right \}$ ($n=2,3$), are found to be correlated on an event by event basis to the initial condition geometrical parameters $\left\{\varepsilon_{2,n}, \Phi_{2,n} \right \}$, respectively. More precisely, we have found that there is a considerable linear correlation between $v_n$ and $\varepsilon_{2,n}$ ($n=2,3$), for three different $p_T$ bins, both in central and peripheral collisions. In addition, we have shown that the phase that defines each flow component, $\Psi_n$, is partially aligned with the respective reference angle $\Phi_{2,n}$, for three different $p_T$ bins (when $0<p_T<3$ GeV), both in central and peripheral collisions. We found that $\Psi_4$ remains generally aligned to $\Phi_{2,4}$ even though $v_4$ and $\varepsilon_{2,4}$ are not strongly linearly correlated. These results show that, considering low and moderate $p_T$, the nonlinear hydrodynamic evolution indeed preserves the global geometric parameters that characterize the initial conditions event by event. In the case of high transverse momentum ($3<p_T<5GeV$) the $\delta_2$ distribution computed in the $\left(0-5 \right)\%$ centrality window is considerably broad.

The phase difference between trigger and associated hadrons, $\Delta_n=\Psi_n^t- \Psi_n^a$, which is in principle nonzero when both angles are defined using different $p_T$ bins, can affect the di-hadron angular correlation function. The remarkable case occurs in the $\left(0-5 \right)\%$ centrality window, when the triggers are taken in the interval $3<p_T<5GeV$ and the associated particles in the interval $0<p_T<1GeV$. Once the associated particles are aligned with the reference angle $\Phi_{2,n}$ and the triggers are not, the factor $\left< \cos \left(2 \Delta_2  \right) \right>$ is rather smaller than the unit ($\sim 0.456$). Consequently, according to the Eq. (\ref{eq:cn}), this result reduces the the contribution of the second harmonic to the di-hadron correlation function. Moreover, we have found (after 1000 events) that the final angular correlation function, $C\left(\Delta \phi \right)$, and its background subtracted version, $R\left(\Delta \phi \right)$, are even functions of $\Delta \phi$. This is a consequence of the parity property of the relative phase distributions $\Delta_n$.

R.~P.~G.~Andrade and J.~Noronha acknowledge
Funda\c c\~ao de Amparo \`{a} Pesquisa do Estado de S\~{a}o Paulo
(FAPESP) and Conselho Nacional de Desenvolvimento Cient\'{\i}fico e Tecnol%
\'{o}gico (CNPq) for financial support. The authors
thank G.~S.~Denicol, Y.~Hama, F.~Grassi, F.~Gardim, M.~Luzum, and J.-Y.~Ollitrault for discussions on the hydrodynamic description of di-hadron angular correlations and A.~Dumitru for providing the Monte Carlo Glauber model.

\appendix

\section{\label{sec:sph} The SPH method}

The SPH formulation of relativistic inviscid hydrodynamics can be done in terms of the variational principle \cite{Elze:1999kc,Aguiar:2000hw}. For the sake of completeness, we shall review this formulation below. We start with the Lagrangian formulation of the relativistic hydrodynamics, in the approximation of an ideal fluid, for vanishing baryon chemical potential. Such a formulation is done by the action

\begin{equation}
I=- \int \epsilon \left[s \left(\vec{x},x^0  \right) \right] \sqrt{-g} d^4x,
\label{eq:apsph1}
\end{equation}

\noindent
under the constraints

\begin{equation}
\left(s u^{\nu} \right)_{; \nu}=\frac{1}{\sqrt{-g}} \partial_{\nu} \left(\sqrt{-g} s u^{\nu} \right)=0
\label{eq:apsph2}
\end{equation}

\noindent
and

\begin{equation}
u^{\mu} u_{\mu}=1,
\label{eq:apsph3}
\end{equation}

\noindent
where $\epsilon$ and $s$ are the energy density and the entropy density of the fluid, respectively (in the local frame); $x=\left(x^0,x^1,x^2,x^3 \right)$ is the generalized coordinate, $g$ is the determinant of the metric tensor, $g_{\mu \nu}$, and $\sqrt{-g}$ is the Jacobian determinant. We shall consider only metrics of the following form

\begin{equation}
\left(g_{\mu \nu} \right) = 
\begin{bmatrix}
g_{00} & 0  \\
0 & -\mathbf{g} 
\end{bmatrix},
\label{eq:apsph4}
\end{equation}

\noindent
where $\mathbf{g}$ is the spatial part of the metric tensor (a 3 x 3 matrix).

By using the formulas (\ref{eq:apsph3}) and (\ref{eq:apsph4}), one finds that

\begin{equation}
\gamma = \frac{1}{\sqrt{g_{00} - \left[\vec{v} \right]^{\mathbf{T} } \mathbf{g}  \vec{v}}}
\label{eq:apsph5}
\end{equation}

\noindent
where $v^i = u^i/u^0$.

Having depicted the Lagrangian formalism for the ideal relativistic hydrodynamics, let us introduce the concepts of the SPH method. 
The basic idea of this method is to parametrize the density of the extensive thermodynamic quantities, each density associated to a conserved charge, in the following way

\begin{equation}
a^* \left(\vec{x},x^0 \right)=\sum_{j=1}^N \nu_j W \left(\vec{x} - \vec{x}_j \left(x^0 \right); h \right),
\label{eq:apsph6}
\end{equation}

\noindent
where the density $a^*$ is associated with an extensive quantity $\mathcal{A}$ and is measured in a space-fixed (calculational) frame. The weights $\nu_j$ are defined by the initial conditions and their values are kept constant during the hydrodynamic expansion. $W$ is a positive-definite function called the kernel function defined using a length scale $h$ called the SPH smoothing parameter and this function has the following properties

\begin{equation}
W \left(\vec{x} - \vec{x}_j \left(x^0 \right); h \right) = W \left(\vec{x}_j  \left(x^0 \right) - \vec{x}; h \right),
\label{eq:apsph7}
\end{equation}

\begin{equation}
\int W \left(\vec{x} - \vec{x}_j \left(x^0 \right); h \right)  d^3 \vec{x} = 1,
\label{eq:apsph8}
\end{equation}

\noindent
and

\begin{equation}
\lim_{h \to 0} W \left(\vec{x} - \vec{x}_j \left(x^0 \right); h \right) = \delta^{3} \left(\vec{x} - \vec{x}_j \left(x^0 \right) \right).
\label{eq:apsph9}
\end{equation}

\noindent
Usually, the coordinates $\vec{x}_j$, which explicitly depend on the ``time" $x^0$, are called SPH Lagrangian coordinates or simply SPH particles. Each one carries a portion $\nu_j$ of the extensive quantity $\mathcal{A}$.

Considering that

\begin{equation}
a^* \to s^* = \sqrt{-g} \gamma s
\label{eq:apsph10}
\end{equation}

\noindent
and

\begin{equation}
\vec{v} \left(\vec{x},x^0 \right) \equiv \frac{1}{s^*} \sum_{j=1}^N \nu_j \frac{d \vec{x}}{d x^0} W \left(\vec{x} - \vec{x}_j \left(x^0 \right); h \right),
\label{eq:apsph11}
\end{equation}

\noindent
it is not difficult to realize that the parametrization (\ref{eq:apsph6}) satisfies the constraints (\ref{eq:apsph2}) and (\ref{eq:apsph3}), independently on the motion of the SPH particles. This is one of the advantages of this method: the entropy is automatically conserved throughout the whole time evolution. Thus, the equation of motion for each one of the SPH particles is obtained from the condition $\delta I=0$.

It is convenient at this point to define the following notation

\begin{equation}
s_i^* = \left(\sqrt{-g}\right)_i \gamma_i s_i.
\label{eq:apsph10a}
\end{equation}

\noindent
The subscript index indicates that the physical quantity must be computed at the position of the i-th SPH particle, i.e., $\vec{x}=\vec{x}_i$. Keeping this notation in mind, the energy density profile can be parameterized as follows:

\begin{equation}
\epsilon \left(\vec{x},x^0 \right)=\sum_{j=1}^N  \epsilon_j  \left(\frac{\nu_j}{s_j^*}\right)  W \left(\vec{x} - \vec{x}_j \left(x^0 \right); h \right)\,.
\label{eq:apsph12}
\end{equation}

\noindent
The quantity $V^* = \left(\frac{\nu_j}{s_j^*}\right)$ is usually called the SPH particle volume. Observe that the extensive thermodynamic quantity used to define the SPH particle volume cannot vanish. This makes the entropy a convenient choice in the case of an ideal fluid. In the case where viscous effects are included and, hence, there is entropy production, a different conserved quantity is chosen to define the SPH particle volume \cite{Denicol:2009am,Noronha-Hostler:2013gga}. 

Introducing the parametrization (\ref{eq:apsph12}) in the action (\ref{eq:apsph1}) one finds that

\begin{equation}
I \to I_{\mathrm{SPH}} = - \sum_{j=1}^N \int \frac{E_j}{\gamma_j} d \tau,
\label{eq:apsph13}
\end{equation}

\noindent
where $E_j=\epsilon_j V_j$ and $V_j=\left(\frac{\nu_j}{s_j}\right)$ is the proper volume of the SPH particle.

Taking into account that $\delta E_j = -p_j \delta V_j$, where $p_j$ is the pressure of the fluid, the condition $\delta I_{\mathrm{SPH}}=0$ leads to the following set of ordinary differential equations, in the hyperbolic coordinate system

\begin{equation}
\frac{d}{d \tau} {\vec{\pi}_{T} \choose \vec{\pi}_{\eta}}_i =  - \sum_{j=1}^N \frac{\nu_i \nu_j}{\tau} \left(\frac{p_i}{\left(\gamma_i s_i\right)^2} + \frac{p_j}{\left(\gamma_j s_j\right)^2} \right) {\nabla_{T} \choose \partial_{\eta}}_i W_{ij},
\label{eq:apsph14}
\end{equation}

\noindent
where

\begin{equation}
\left( \vec{\pi}_{T} \right)_i = \nu_i \gamma_i \left(\frac{\epsilon_i + p_i}{s_i} \right) \left(\vec{v}_T \right)_i,
\label{eq:apsph15}
\end{equation}

\begin{equation}
\left( \vec{\pi}_{\eta} \right)_i = \tau^2 \nu_i \gamma_i \left(\frac{\epsilon_i + p_i}{s_i} \right) \left( v_{\eta} \right)_i,
\label{eq:apsph16}
\end{equation}

\begin{equation}
\gamma_i = \left(1 - \left(\vec{v}_T \right)_i^2 - \tau^2 \left( v_{\eta} \right)_i^2  \right)^{-\frac{1}{2}},
\label{eq:apsph17}
\end{equation}

\begin{equation}
\left(\nabla_{T} \right)_i = \left(\partial_x,\partial_y \right)_i
\label{eq:apsph18}
\end{equation}

\noindent
and

\begin{equation}
W_{ij} = W\left(\left(\vec{r}_T \right)_i - \left(\vec{r}_T \right)_j, \eta_i - \eta_j; h \right).
\label{eq:apsph19}
\end{equation}

\noindent
In Eqs.\ (\ref{eq:apsph14} - \ref{eq:apsph19}): $\left(\vec{v}_T\right)_i = \left(d \vec{r}_T/d \tau\right)_i$, $\left(v_{\eta} \right)_i=\left(d \eta/ d \tau \right)_i$, $\tau=\sqrt{t^2 - z^2}$, $\eta=0.5 \ln \left[\left(t+z \right)/ \left(t-z \right) \right]$ and $\vec{r}_T = \left(x,y \right)$.

Thus, one can see that a feature of the SPH approach is that the dynamics of the relativistic fluid is described by a set of ordinary differential equations whose solutions can be obtained by simple numerical methods.

In the boost-invariant Ansatz, it is not difficult to realize that the longitudinal equation is trivially satisfied, once $v_{\eta}=0$ and the pressure gradient vanishes along the $\eta$ direction. Moreover, in this solution

\begin{equation}
W\left(\left(\vec{r}_T \right)_i - \left(\vec{r}_T \right)_j, \eta_i - \eta_j; h \right) \to W\left(\left(\vec{r}_T \right)_i - \left(\vec{r}_T \right)_j; h \right),
\label{eq:apsph20}
\end{equation}

\noindent
since the transverse flow is identical in any transverse plane. 

\subsection*{\label{subsec:npar} Numerical parameters}

In the SPH method there are three basic parameters: the width of the $W$ function, $h$, the total number of SPH particles, $N$, and the size of the time step $d \tau$ used in the numerical solution of the ordinary differential equations (\ref{eq:apsph14}), that determine the dynamics of each SPH particle. The parameter $h$ fixes the resolution of the interpolation formula (\ref{eq:apsph6}), i.e., the smaller is $h$, more detailed the profile of the density $a^*$ is. The parameter $N$, taking into account the fact that the SPH particles move together with the fluid, must be large enough to guarantee a minimal number of SPH particles inside an arbitrary area $\delta A \approx \pi h^2$ (in the case of a three dimensional calculation, an arbitrary volume should be considered). In other words: for a fixed $h$, the hydrodynamic solution should not depend on the parameter $N$. In general, for a given $h$ one increases $N$ until the quantities computed become insensitive to further changes in this parameter.

In this work, it is used $h=0.3$fm, $N\sim 70000$, and $d \tau=0.1$fm. This choice for $h$ preserves all the interesting structure present in the initial conditions and this value of $N$ is large enough to guarantee convergence of the results computed in this paper. By using these parameters, the relative error in the total energy conservation, comparing the energy at beginning of the simulation with the energy at the end, is smaller than $0.1\%$. For two consecutive steps, the relative error is smaller than $0.001\%$. 

\section{\label{sec:gubser} Gubser flow}

The analytical solution obtained by Gubser for the transverse flow (azimuthally symmetric) of a boost invariant, ideal and conformal ($\epsilon =3p$) fluid is the following \cite{Gubser:2010ze}:

\begin{equation}
\epsilon \left(\tau,r \right)= \frac{\epsilon_0 \left(2q \right)^\frac{8}{3}}{\tau^\frac{4}{3} \left(1+2q^2 \left(\tau^2+r^2 \right)+ q^4 \left(\tau^2 - r^2 \right)^2   \right)^\frac{4}{3}}
\label{eq:apgub01}
\end{equation}

\noindent
and

\begin{equation}
v_T\left(\tau,r \right)= \tanh \left[ \kappa \left(\tau,r \right) \right]=\frac{2q^2 \tau r}{1+q^2 \tau^2 + q^2 r^2},
\label{eq:apgub02}
\end{equation}

\noindent
where $v_T$ is the transverse velocity, $q$ is an arbitrary parameter (with dimension fm$^{-1}$), which is related to the transverse distribution of matter,  and $\epsilon_0$ is a dimensionless normalization factor.

\begin{figure}
\includegraphics[scale=0.40]{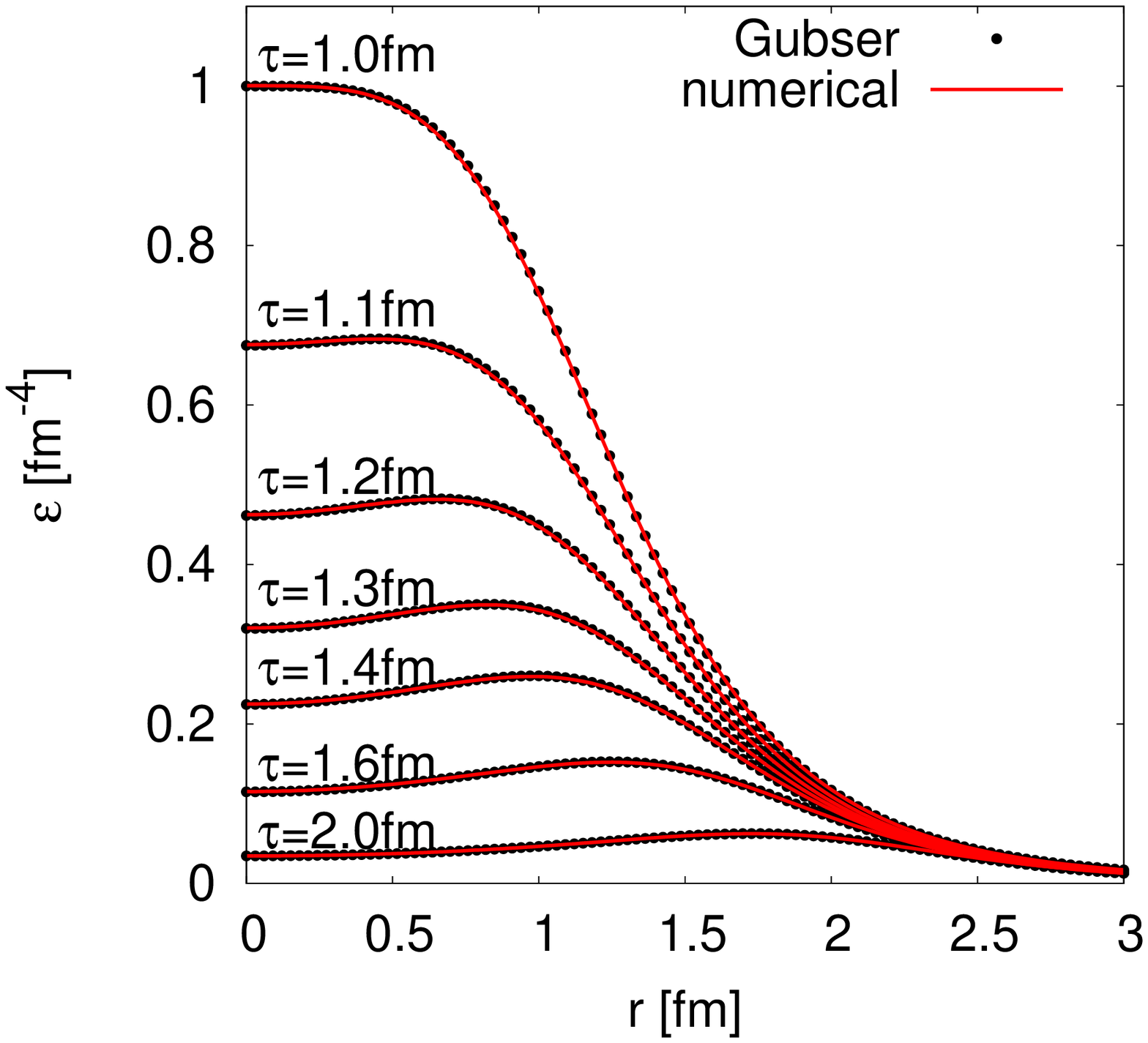}% Here is how to import EPS art
\\
\includegraphics[scale=0.40]{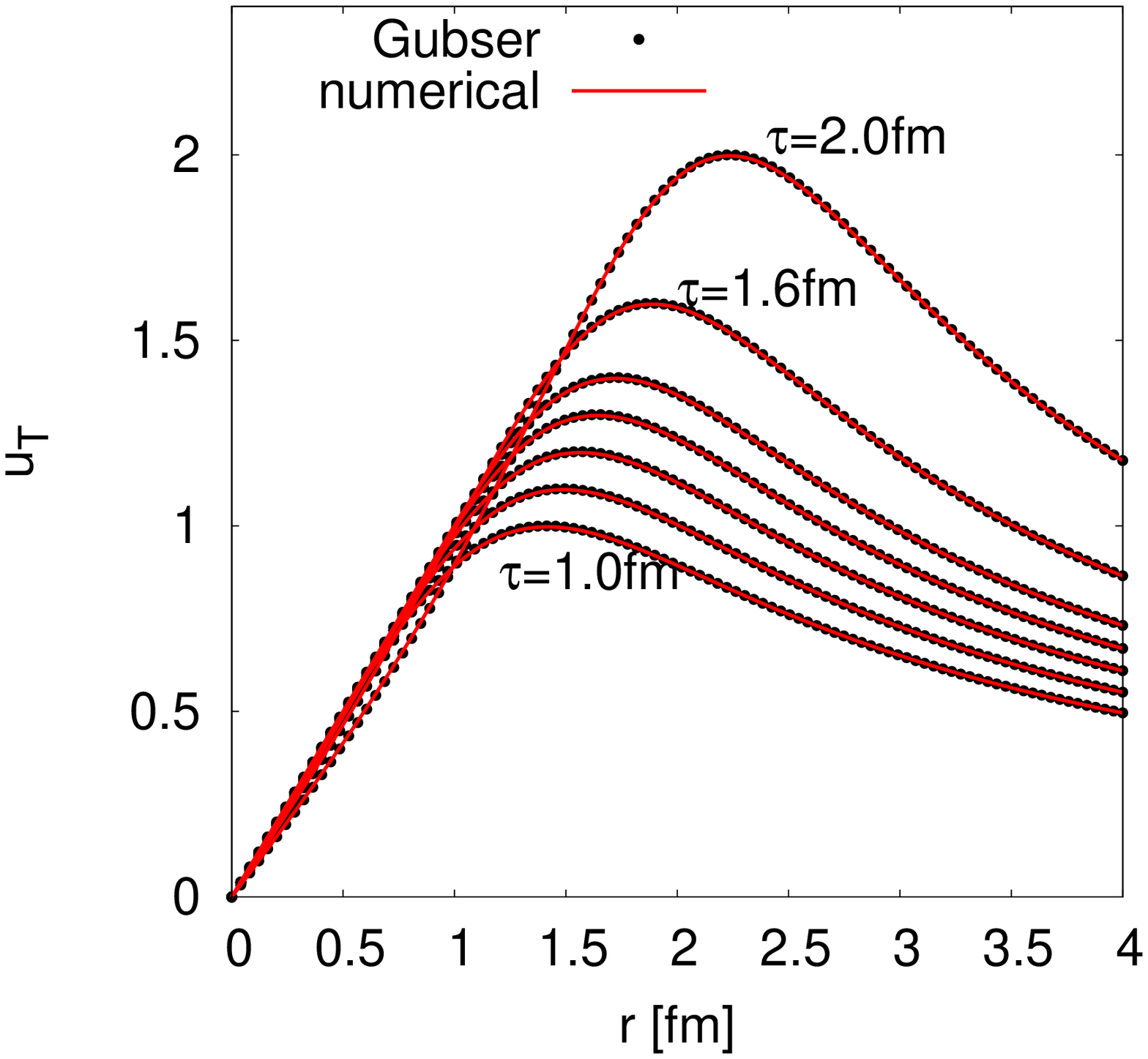}% Here is how to import EPS art
\caption{\label{fig:gubser2} Energy density distribution and transverse 4-velocity ($u_T=\sinh \kappa$) defined as a function of the transverse radius $r$. The dots represent the exact solution and the solid lines the numerical computation. We used $\epsilon_0 = 1$ and $q=1$fm$^{-1}$.}
\end{figure}

In Fig.\ \ref{fig:gubser2} we show the comparison between the exact solution and our numerical computation for the energy density distribution and transverse 4-velocity ($u_T = \sinh \kappa$), both quantities as functions of the transverse radius coordinate $r$. For the sake of simplicity, it is used $\epsilon_0 = 1$ and $q=1$fm$^{-1}$. As one can see, the code reproduces the exact solution with great accuracy.

\section{\label{sec:cfpsph} Cooper-Frye Prescription in the SPH approach}

In the Cooper-Frye prescription \cite{Cooper:1974mv} the invariant momentum distribution is given by

\begin{equation}
E \frac{dN}{d^3 \vec{p}}= \frac{g}{\left(2 \pi \right)^3} \int_{\Sigma} f \left(T_{fo},p^{\nu}u_{\nu} \right)  p^{\mu} d \sigma_{\mu},
\label{eq:apcf01}
\end{equation}

\noindent
where $f$ is the particle distribution as a function of the momentum $\vec{p}$. We have assumed zero baryon chemical potential. The integral is done on the hypersurface $\Sigma$, characterized by a  constant temperature (the freeze-out temperature $T_{fo}$). In the case of ideal hydrodynamics, $f$ is the thermal equilibrium distribution.

\noindent
In the SPH representation (see the appendix \ref{sec:sph}), the Cooper-Frye formula (\ref{eq:apcf01}) becomes \cite{Osada:2001hw}

\begin{equation}
E \frac{dN}{d^3 \vec{p}}= \sum_{j=1}^N  p^{\nu} \left(q_{\nu} \right)_j f_j,
\label{eq:apcf02}
\end{equation}

\noindent
where

\begin{equation}
\left(q_{\nu} \right)_j = \frac{g}{\left(2 \pi \right)^3} \frac{\left(n_{\nu} \right)_j}{\left| \left(n_{\mu} \right)_j \left(u^{\mu} \right)_j  \right|} \frac{\nu_j}{s_j},
\label{eq:apcf03}
\end{equation}

\noindent
and

\begin{equation}
f_j = \frac{1}{\exp \left[p^{\nu} \left(u_{\nu} \right)_j /T_{fo} \right] \pm 1}.
\label{eq:apcf04}
\end{equation}

\noindent
In the formula (\ref{eq:apcf02}), the summation is over all SPH particles. The quantities with the index $j$ must be computed at the position of the j-th SPH particle when it achieves the freeze-out hypersurface $\Sigma$. The quantity $\left(n_{\nu} \right)_j$ is the  normal to this hypersurface.

The formula (\ref{eq:apcf02}) in the hyperbolic coordinate system can be put in the following form

\begin{equation}
\frac{dN}{dp^x dp^y dp^{\eta}}= \sum_{j=1}^N f_j \left[ \tau \left(q_{\tau} \right)_j  -  \tau \frac{\vec{p}_{T} \cdot \left( \vec{q}_{T} \right)_j}{p^{\tau}} \right],
\label{eq:apcf05}
\end{equation}

\noindent
where

\begin{equation}
f_j = \frac{1}{\Lambda_j \exp \left[p^{\tau} \left(u^{\tau} \right)_j /T_{fo} \right] \pm 1},
\label{eq:apcf06}
\end{equation}

\noindent
$\vec{p}_T = \left(p^x,p^y \right)$ and $\Lambda_j = e^{-\vec{p}_{T} \cdot \left( \vec{u}_{T} \right)_j/T_{fo}}$. In the Eq.\ (\ref{eq:apcf05}) it was used that $\eta=0$ (the computation is done at the mid-rapidity) and $\left(q_{\eta}\right)_j=0$ (in the boost invariant solution the normal to the hypersurface $\Sigma$ does not have a longitudinal component). In Eq.\ (\ref{eq:apcf06}) we used that $u^{\eta}=0$ (longitudinal boost invariant flow).

The distribution of particles as a function of the transverse momentum can be obtained by the integration of Eq.\ (\ref{eq:apcf05}) with respect to the longitudinal momentum $p^{\eta}$. In this kind of calculation, the SPH particles are placed only in one transverse plane: the mid-rapidity transverse plane, once the transverse hydrodynamic evolution is boot invariant. This integration takes into account this symmetry. Thus,

\begin{equation}
\frac{dN}{dp^x dp^y}= \sum_{j=1}^N \left[\left(q_{\tau} \right)_j m_{T} \left(I_2^{\pm} \right)_j  -   \vec{p}_{T} \cdot \left( \vec{q}_{T} \right)_j  \left(I_1^{\pm} \right)_j \right]
\label{eq:apcf07}
\end{equation}

\noindent
where

\begin{equation}
\left(I_{1}^{\pm}\right)_j= \int_{-\infty}^{\infty} \frac{ \left( 1 + x^2 \right)^{-\frac{1}{2}}}{\Lambda_j  \exp \left[\sigma_j \sqrt{1+ x^2} \right] \pm 1} dx,
\label{eq:apcf08}
\end{equation}

\begin{equation}
\left(I_{2}^{\pm}\right)_j= \int_{-\infty}^{\infty} \frac{1}{\Lambda_j  \exp \left[\sigma_j \sqrt{1+ x^2} \right] \pm 1} dx,
\label{eq:apcf09}
\end{equation}

\noindent
$\left(\vec{q}_T\right)_j=\left(q^x,q^y \right)_j$, $m_{T}=\sqrt{\vec{p}_{T} \cdot \vec{p}_{T} + m^2}$ is the transverse mass and $\sigma_j = m_T \left(u^{\tau} \right)_j/T_{fo}$\,.

By using the integral form of the modified Bessel functions $K_{\nu}$, the integrals (\ref{eq:apcf08}) and (\ref{eq:apcf09}) can be written
as

\begin{equation}
\left(I_{1}^{+}\right)_j= \sum_{n=1}^{\infty} \left(-1 \right)^{n+1} \frac{2}{\left(\Lambda_j\right)^n}K_0 \left(n\sigma_j \right),
\label{eq:apcf10}
\end{equation}

\begin{equation}
\left(I_{1}^{-}\right)_j= \sum_{n=1}^{\infty}  \frac{2}{\left(\Lambda_j\right)^n}K_0 \left(n\sigma_j \right), \hspace{1.3cm}
\label{eq:apcf11}
\end{equation}

\begin{equation}
\left(I_{2}^{+}\right)_j= \sum_{n=1}^{\infty} \left(-1 \right)^{n+1} \frac{2}{\left(\Lambda_j\right)^n}K_1 \left(n\sigma_j \right),
\label{eq:apcf12}
\end{equation}

\noindent
and

\begin{equation}
\left(I_{2}^{-}\right)_j= \sum_{n=1}^{\infty}  \frac{2}{\left(\Lambda_j\right)^n}K_1 \left(n\sigma_j \right)\,. \hspace{1.3cm}
\label{eq:apcf13}
\end{equation}

Finally, it is important to mention that in this approach there is an interval around mid-rapidity which is implicitly defined when the entropy portion $\nu_j$ is assigned to the SPH particles at the beginning of the simulation (see Eqs. \ref{eq:apsph6} and \ref{eq:apsph10}).

\nocite{*}

\bibliography{draftv2}% Produces the bibliography via BibTeX.

\end{document}